\documentstyle[prd,aps]{revtex}
\tighten
\def\be{\begin{equation}}
\def\ee{\end{equation}}
\def\beq{\begin{eqnarray}}
\def\eeq{\end{eqnarray}}

\begin{document}
\begin{titlepage}
\begin{flushright}  
\end{flushright}
\begin{center}
   \vskip 2em
  {\bf \LARGE GENERALIZED CHIRAL MEMBRANE} 
  \vskip 2em
 {\bf \LARGE  DYNAMICS}
  \vskip 2em
{\large Rub\'en Cordero${}^{(1)}$ and Efra\'{\i}n 
Rojas${}^{(2)}$ \\[1em]}
\em{
${}^{(1)}$Departamento de F\'{\i}sica, \\
Escuela Superior de F\'{\i}sica y Matem\'aticas 
del IPN\\
Edificio 9, 07738, M\'exico D.F., MEXICO}\\[1em]
${}^{(2)}$ Departamento de F\'{\i}sica \\
Centro de Investigaci\'on
y de Estudios Avanzados del IPN \\
Apdo Postal 14-740, 07000 M\'exico,
D. F.,
MEXICO \\[1em]

\end{center}
\vskip 1em
\begin{abstract}
We develop the dynamics of the chiral superconducting 
membranes (with null current) in an alternative 
geometrical approach. Besides of this, we show the 
equivalence of the resulting description to the one 
known Dirac-Nambu-Goto 
(DNG) case. Integrability for chiral string model is 
obtained using a proposed light-cone gauge. In a similar 
way, domain walls are integrated by means of a simple 
ansatz. 
\end{abstract}

\begin{center}

PACS: 98.80.Cq, 98.80Hw, 11.27+d

\end{center}

\end{titlepage}

\section {Introduction}

It is believed that cosmic strings are fundamental 
bridges in the understanding of the Universe formation
due to that several cosmological phenomena can be 
described by means of the cosmic strings properties.
Besides of these there are other kinds of cosmic 
objects possesing different properties of those 
inherited to ordinary cosmic strings, for example:
domain walls, hybrid structures like domain walls
bounded by strings and so forth, \cite{Vilenkin1}. 
They can arise in several Grand Unified Theories  
whenever exist an appropriate symmetry breaking scheme. 
However, there is other class of cosmological objects that
can emerge with the ability to carry some sort of charge.  
For instance, as was suggested by Witten \cite{Witten} 
in the middle of the eighties, cosmic strings could 
behave like superconductors.

Since that time the vast research on superconducting
strings has thrown a new variety of cosmic objects.
The cosmology resulting of supersymmetric theories
(SUSY) was also considered yielding to other class of 
cosmic strings, namely chiral cosmic strings. These 
objects are the result of a symmetry breaking in SUSY
where a $U(1)$ symmetry is broken with a Fayet-Iliopoulos $D$ 
term, turning out a sole fermion zero mode traveling in
only one direction in the string core \cite{Davis2}.
In other words, when the current along the superconducting 
string shows a light-like causal structure then we 
have a chiral string. Carter and Peter \cite{Carter}
have made an exhaustive study of this kind of cosmic strings
and a few time later on it has been continued by other authors 
\cite{Vilenkin,Davis1}. 

The purpose of this contribution 
is extend in an alternative geometrical 
way the dynamical results for chiral strings reported 
in \cite{Carter,Vilenkin} using a Kaluza-Klein (KK)
reduction mechanism \cite{Nielsen} and following
closely the variational techniques developed in 
\cite{Defo}.
Bearing in mind the KK idea, and assuming our original 
background spacetime to be 4-dimensional, the 
generalization to higher dimensional objects (membranes)
tracing out worldsheets is possible. From this assumption 
we found that the dynamics of the chiral membranes 
resemble to a 5-dimensional DNG case. We describe now our
membrane with an extended embedding. This description
has the advantadge of treat the new membrane on the 
same footing as an ordinary DNG membrane.
It is often the case that while an existing 
theory admits a number of equivalent descriptions, 
one of them suggets generalizations and simplicities 
more readily than others. It is our goal.

The equations of motion are a generalization of those 
of DNG type, i.e., the motion of the chiral membranes
looks like minimal surfaces in a KK space but subject 
to a particular condition. In addition, we 
have the current conservation on the membrane which 
emerges as the remain equation of motion for the extra 
variable.

The note is organized as follows: In section 2 we 
develop the essential mathematical features to describe 
the superconducting chiral membranes. This is done by 
exploting the theory of deformations achieved in 
\cite{Defo}. 
In section 3, we present a simplified 
version for the dynamics of chiral extended objects, 
in contrast with others approachs. In section 4 we 
specialize to the chiral string model. We found
a new method of integrability for chiral string model using a 
light-cone gauge. In section 5 we found integrability
for a simple chiral domain wall model.
Finally, we give some conclusions.

\section{Geometry for Chiral Membranes}

In this section we describe both the intrinsic and
extrinsic geometry for chiral membranes, i.e.,
possesing null currents on the worldsheet ($\omega =
\gamma^{ab}\, \phi_{,a} \phi_{,b} = 0$), based in the 
Kaluza-Klein approach achieved in \cite{Nielsen}. The
present development is close to the conceptual framework
made in \cite{Defo}. 
To begin with, we consider a relativistic membrane 
of dimension $d$, whose 
worldsheet $\{ m, \Gamma_{ab} \}$ is an oriented 
timelike $d+1$-dimensional manifold, embedded in a 
5-dimensional extended arbitrary fixed background 
spacetime $\{ {\cal M}, g_{{\bar{\mu} \bar{\nu}}} \}$,
$\bar{\mu}= 0,1,...,4$. We shall describe
the worldsheet by the extended embedding
\begin{equation}
X^{\bar{\mu}}= 
\left(
\begin{array}{l}
X^\mu (\xi^a)\\
\,\,\,\phi (\xi^a)
\end{array}
\right)\,\,,
\label{eq:extembb}
\end{equation}
where $\phi$ is a field living on the worldsheet $m$; 
$a,b= 0,1,2$, and $\xi^a$ are coordinates on the 
worldsheet. With the former 
embedding, we can make contact with the 
Kaluza-Klein description for 
the background space-time metric
\begin{equation}
g_{\bar{\mu} \bar{\nu}}= 
\left(
\begin{array}{ll}
g_{\mu \nu}&0 \\
0&g_{44} 
\end{array}
\right)\,\,,
\label{eq:newmetric}
\end{equation}
where $g_{\mu \nu}$ is the metric of the original background 
spacetime, and $g_{44}$ is a constant. 
The tangent basis for the worldsheet is defined by
$
e_a := X^{\bar{\mu}}{}_{,a} \partial_{\bar{\mu}} 
= e^{\bar{\mu}}{}_a \partial_{\bar{\mu}} 
$
where the prime denotes partial derivative with
respect to the coordinates $\xi^a$. The tangent 
vectors $e^{\bar{\mu}}{}_a$, associated with the 
embedding (\ref{eq:extembb}), can be written as
\begin{equation}
e^{\bar{\mu}}{}_a= 
\left(
\begin{array}{l}
e^\mu {}_a\\
\phi_{,a}
\end{array}
\right)\,\,.
\label{eq:extvect}
\end{equation}
The metric induced on $m$ is given by
$
\Gamma_{ab} = g_{\bar{\mu} \bar{\nu}}
e^{\bar{\mu}}{}_a e^{\bar{\nu}}{}_b \nonumber =  
\gamma_{ab} + g_{44} \phi_{,a}\phi_{,b}\,\,,
$
where $\gamma_{ab}= g_{\mu \nu}e^\mu {}_a e^\nu {}_b$
is the standard metric for the worldsheet without
the field $\phi$. 
The normal basis for the worldsheet is denoted
by $n^{\bar{\mu}\,I}$ which is intrinsically 
defined by 
\begin{equation}
g_{\bar{\mu} \bar{\nu}}n^{\bar{\mu}\,I} 
n^{\bar{\nu}\,J}
= \delta^{IJ}\,, \hspace{1cm} g_{\bar{\mu} 
\bar{\nu}}n^{\bar{\mu}\,I} e^{\bar{\nu}}{}_a 
= 0\,,
\end{equation}
where $I,J = 1,2,...,N - d$. We can write explicitly
the complete orthonormal basis, which we label as
$n^{\bar{\mu}\,I} =
\{ n^{\bar{\mu}\,i} , n^{\bar{\mu}\,(4)}\}$ as 
follows,
\begin{equation}
n^{\bar{\mu}\,i}=
\left(
\begin{array}{l}
n^{\mu \,i}\\
0
\end{array}
\right)\,\,,
\hspace{2cm}
n^{\bar{\mu}\,(4)}=
\sqrt{g_{44}} 
\left(
\begin{array}{l}
e^\mu {}_a \phi^{,a} \\
- g^{44}
\end{array}
\right)\,\,,
\label{eq:extvect}
\end{equation}
where we have assumed that $n^{\mu \,i}$ satisfy
$g_{\mu \nu} n^{\mu \,i}n^{\nu \,j}= \delta^{ij}$, and
$i$ take the values $i= 1,...,4-d$.
One of the more important quantities determining the 
extrinsic geometry is 
the extrinsic curvature $K_{ab} ^I$ along the normal 
basis is defined as
\begin{equation}
K_{ab} ^I = - n_{\bar{\mu}} {}^I D_a e^{\bar{\mu}}{}_b \,.
\end{equation}
where $D_a := e^{\bar{\mu}}{}_a D_{\bar{\mu}} $ is 
the gradient along the tangential basis, and 
$ D_{\bar{\mu}} $ is the covariant derivative 
compatible with $g_{\bar{\mu}\bar{\nu}}$.
The last expression can be splitted as follows: {\it i})
For $I=i$ we have, $K_{ab} ^i= - n_\mu {}^i 
D_a e^{\mu}{}_b$ which is the well known expression
for the extrinsic curvature for the worldsheet of the 
membrane \cite{Defo},
and {\it ii}) for $I=4$, $K_{ab} ^{(4)}
= \sqrt{g_{44}} \,\nabla_a \nabla_b \phi$, where 
$\nabla_a$ is the covariant derivative compatible 
with $\gamma_{ab}$. The index $(4)$ denotes the direction
along the normal $n^{\bar{\mu}\,(4)}$.

\vspace{0.4cm}

\section{Chiral Membrane Dynamics}

In this section we will show the equivalence between the
chiral membrane dynamics and the DNG dynamics in an 
extended background spacetime plus a chirality condition. 
The starting point to discuss the dynamics of chiral 
membranes is the DNG like action which is invariant under 
reparametrizations of the worldsheet $m$,
\begin{equation}
S= - \mu_0 \,\int_m d^{d+1} \xi \,\sqrt{- \Gamma} \,\,,
\label{eq:chiaction}
\end{equation}
where $\Gamma$ is the determinant of the induced metric
(\ref{eq:newmetric}) from the spacetime by the embedding 
(\ref{eq:extembb}), and $\mu_0$ is a constant.
The determinant is straightforward computed and given by
$\Gamma = \gamma \,(1 + g_{44}\,\omega)$, where $\gamma$ 
is the determinant of the old induced metric on the 
worldsheet, $\gamma_{ab}$. The action (\ref{eq:chiaction})
turn into
\begin{equation}
S= - \mu_0 \,\int_m d^{d+1} \xi \sqrt{- \gamma}\,
(1 + g_{44}\,\omega)^{1/2} \,.
\label{eq:cuasi-act}
\end{equation}
Observe that the resulting action from the DNG 
like action (\ref{eq:chiaction}),
is the one for superconducting strings involving 
the Nielsen model, where ${\cal L}(\omega)= 
\sqrt{1 + g_{44}\,\omega}$, \cite{Nielsen}.
In other words, the superconducting string theory
with the Nielsen model is equivalent to DNG like 
action (\ref{eq:chiaction}). 

An important issue that deserve attention is 
that of the equations of motion which are already 
known \cite{Carter,Vilenkin}. We want rebound 
here the geometrical framework introduced in 
section 2 in the attainment of chiral membrane dynamics.
Using similar variational techniques to that
developed in \cite{Defo} we can get 
the equations of motion from the action 
(\ref{eq:chiaction}). It is worthy to mention that
this method is very graceful because rebound the 
geometrical nature of the worldsheet. The variation
of the action (\ref{eq:chiaction}) gives
\begin{eqnarray}
\delta S &=& -\mu_0 \int_m d^{d+1} \xi \,\frac{1}{2}\,
\sqrt{ -\Gamma}\,\Gamma^{ab} \,\delta \Gamma_{ab} 
\nonumber =  -\mu_0 \int_m d^{d+1} \xi \,\sqrt{ -\Gamma}\, 
\Gamma^{ab} K_{ab} ^I \Phi_I = 0\,\,,
\end{eqnarray}
where we have considered only normal deformations
to the worldsheet\footnote{The tangential deformations 
can be identified with the actions of worldsheet 
diffeomorphisms  so we can ignore them since we are 
interested in quantities invariant under reparameterizations 
of the worldsheet. These tangential deformations are important 
in the study of composite objects \cite{Defo,Cordero4}.}, 
$\Phi^I$ are the deformation
normal vector fields and $\Gamma^{ab}$ is the 
inverse metric of $\Gamma_{ab}$ given by $\Gamma^{ab}
= \gamma^{ab} - g_{44} \nabla^a \phi \nabla^b \phi$. 
We can immediately read the equations of motion
\begin{equation}
\mu_0 \Gamma^{ab} K_{ab} ^I = 0 \,\,.
\label{eq:eqmotion}
\end{equation}
Let is worth noticing the similarity of these
equations with those ones arising for minimal surfaces, 
namely, $\gamma^{ab} K_{ab} ^i = K^i = 0$, \cite{Defo}.
In fact, in our description $\Gamma^{ab}$ play 
the role of a metric. Let us now decode the 
several cases involve in the Eq. (\ref{eq:eqmotion}).
a) $I=i$. The equations of motion take the form,
\begin{equation}
\mu_0 \gamma^{ab} K_{ab} ^i - \mu_0 g_{44}\nabla^a \phi 
\nabla^b \phi K_{ab} ^i = 0\,.
\label{eq:eqmotion1}
\end{equation}
On other hand, in the generic superconducting 
membranes picture, the strees-energy-momentum tensor 
adquires the form $T_{ab} = {\cal L}(\omega) \gamma_{ab} 
- 2 \,(d {\cal L}/d \omega)\, \nabla_a \phi 
\nabla_b \phi$, where ${\cal L}(\omega)$ is a function
of $\omega$, depending on the 
particular models \cite{Carter2}. 
When the chiral current limit is taken into account, 
the quantities ${\cal L}(\omega)$ and 
${d {\cal L}}/{d \omega}$, adquire constant 
values. If we define 
$g_{44}:= 2(d {\cal L}/d \omega)|_{\omega =0}$
and $\mu_0 = {\cal L}(\omega)|_{\omega =0}$, we can identify 
the Eq. (\ref{eq:eqmotion1})
with the standard equations of motion, namely:
${ T}^{ab} K_{ab} ^i = 0$, \cite{Carter2}.
b) $I=(4)$. In this case we have now directly,
\begin{equation}
\Gamma^{ab} K_{ab} ^{(4)} =0= \nabla_a \nabla^a
\phi \,,
\end{equation}
which is a wave equation for $\phi$, corresponding
to a conserved current carrying onto the 
worldsheet for chiral currents.

\section{Chiral String Model}

We specialize now to the case 
of chiral strings. We illustrate the 
chiral string model from a Lagrangian point of view. 

The presence of the gauge symmetry in a field theory
means that not all of the field components, $X^{\bar{\mu}}
(\xi^a)$, are dynamical. In our case, the 
reparameterization invariance allow us to choose 
a gauge in which the dynamical equations are 
tractable. Due to we have considered a DNG like action, 
(\ref{eq:chiaction}), we have the freedom of choice 
an acceptable gauge condition. Several authors have 
reported solutions for the chiral string model using
specific ansatzs \cite{Carter,Vilenkin,Davis1}. In this 
contribution we get a solution 
for the mentioned model using a different gauge. For a 
review of those solutions using our description, see
the Ref. \cite{ChiralMemb}.

We consider the light-cone 
gauge over the spacetime coordinates adapted to 
our description. It is 
well known that the orthonormal gauge do not 
fully fix the gauge because there is residual 
reparametrization invariance. A favorite gauge 
choice that fix the gauge and allow us to solve 
the constraints is the light-cone gauge.  
This gauge was used by Hoppe in the search for
explicit solutions for the classical equations
of motion of relativistic membranes \cite{Hoppe}.
We shall use this orthonormal light-cone gauge
in the search of integrability for the chiral
string model.

To proceed further,
we assume the original background metric to be flat,
$g_{\mu \nu} = \eta_{\mu \nu}$, with signature
$(-,+,+,+)$ and the embedding (\ref{eq:extembb}).
For the KK spacetime we define light-cone 
coordinates, $X^+$ and $X^-$, as
\begin{eqnarray}
X^+ &=& \frac{1}{\sqrt{2}}\,(X^0 + 
       \sqrt{g_{44}}\,\phi) \,, 
\label{eq:X+}       \\
X^- &=& \frac{1}{\sqrt{2}}\,(X^0 - 
       \sqrt{g_{44}}\,\phi)\,,
\label{eq:X-}      \\      
\vec{X} &=& (X^1,X^2,X^3)\,. 
\end{eqnarray}               
The light-cone gauge points $\tau$ along $X^+$,
\begin{equation}
X^+ = X^+ _0 + P^+ \,\tau \,,
\end{equation}
where $X^+ _0$ and $P^+$ are constants. The 
idea is solve for $X^-$ leaving the $X^i$
variables, where $i=1,2,3$. Laying hold of 
the orthonormal light-cone gauge \cite{Hoppe},
\begin{equation}
\Gamma_{ab}= 
\left(
\begin{array}{ll}
\Gamma_{\tau \tau} & \,\, 0\\
\,0  & \Gamma_{AB}
\end{array}
\right)\,\,,
\label{eq:algo}
\end{equation}
where $\Gamma_{ab}$ is given by 
(\ref{eq:newmetric})
and $A,B=1,...,d$, besides of 
$\sqrt{-\Gamma}\,\Gamma^{\tau \tau} = -1$, we can simplify 
the equations of motion, $\partial_a(\sqrt{-\Gamma}\,
\Gamma^{ab}\,X^{\bar{\mu}}{} _{,b})= 0$ in the set of 
equations
\begin{eqnarray}
{\cal D} \,X^{\bar{\mu}} &=& 0 \,,
\label{eq:lcemotion} 
\\
2 P^+ \dot{X} ^- &=& \dot{\vec{X}}\cdot \dot{\vec{X}}
                    + \bar{\Gamma} \,,
\label{eq:constraint1}                    
\\
P^+ X^- {}_{,A}&=& \dot{\vec{X}}\cdot \vec{X}_{,A}
\label{eq:constraint2}
\end{eqnarray}
where we have defined $\bar{\Gamma}:= {\mbox{det}}
(\Gamma_{AB}) = - \Gamma_{\tau \tau}$ and we have 
defined the differential operator
\begin{equation}
{\cal D} := - \partial_\tau ^2 + \partial_A 
(\bar{\Gamma}\,\Gamma^{AB}\,\partial_B )\,.
\end{equation}
The Eq. (\ref{eq:lcemotion}) represents the 
equations of motion in this gauge, 
(\ref{eq:constraint1}) and (\ref{eq:constraint2})
are the constraints relations for the system.
Deriving with respect to $\tau$ the Eq. 
(\ref{eq:constraint1}), we can rewrite it as
$P^+ {\cal D} X^- = \dot{\vec{X}}\cdot {\cal D}
\vec{X}$; so if
\begin{equation}
{\cal D} \vec{X} =0 \,,
\label{eq:lcemotion1}
\end{equation}
we get the condition ${\cal D} X^- = 0$, 
which we can observe from (\ref{eq:lcemotion}).
Thus, we have reduced the problem to solve the set 
(\ref{eq:constraint1}), (\ref{eq:constraint2}) 
and (\ref{eq:lcemotion1}). So far the results 
are general for minimal surfaces of arbitrary
dimension. Now we specialize to the case of 
chiral strings. In order to get integrability
for the chiral string model, besides of Eqs. 
(\ref{eq:constraint1}), (\ref{eq:constraint2}) 
and (\ref{eq:lcemotion1}), the condition $\omega=0$
must be considered. Using the stringy 
notation\footnote{As is well known, in such 
case the worldsheet is parametrized by the coordinates
$\xi^0 = \tau$ and $\xi^1 = \sigma$. The symbols $^.$ and
$'$ denote partial derivatives with respect to $\xi^0$
and $\xi^1$, respectively.},
the appropiate expressions are
\begin{equation}
\Gamma_{ab}= 
\left(
\begin{array}{ll}
- \vec{X}' \cdot \vec{X}' & \,\,\,\,\,\,\,\,0\\
\,\,\,\,\,\,\,\,0  & \vec{X}' \cdot \vec{X}'
\end{array}
\right)\,\,,
\label{eq:algomas}
\end{equation}
with $\Gamma_{\tau \tau}= - \Gamma_{\sigma \sigma}$.
The equations of motion (\ref{eq:lcemotion1})
transform in the wave equation: ${\cal D} \vec{X}
= (- \partial_\tau ^2 + \partial_\sigma ^2 )\vec{X}
= -\ddot{\vec{X}} +\vec{X}'' =0$, whose general 
solution is given by
\begin{equation}
\vec{X} = \vec{a} (\tau + \sigma) + \vec{b}(\tau -
\sigma) \,.
\end{equation}
The constraints (\ref{eq:constraint1}) and
(\ref{eq:constraint2}) adquire the form
\begin{eqnarray}
P^+ \dot{X}^- &=& |\vec{a}`|^2 + |\vec{b}`|^2 \,,
\label{eq:constraint3} \\
P^+  X^{- '}  &=& |\vec{a}`|^2 - |\vec{b}`|^2 \,,
\label{eq:constraint4}
\end{eqnarray}
where we have used the notation $\vec{a}`$ and 
$\vec{b}`$ to denote derivatives with respect to 
their arguments. From the Eqs. (\ref{eq:X+}) and
(\ref{eq:X-}) as well as Eq. (\ref{eq:newmetric}),
we can separate the metric $\gamma_{ab}$,
\begin{equation}
\gamma_{ab}= 
\left(
\begin{array}{ll}
- \vec{X}' \cdot \vec{X}' - \frac{1}{2}
\,(P^+ - \dot{X} ^-)^2 & \,\,\,\frac{1}{2}
(P^+ - \dot{X} ^-) X^{- '}\\
\,\,\,\,\,\,\,\,\,\,\,\,\frac{1}{2}\,(P^+ - 
\dot{X} ^-) X^{- '}  & 
\vec{X}' \cdot \vec{X}' - \frac{1}{2}\,(X^{-'})^2
\end{array}
\right)\,\,,
\label{eq:algomas1}
\end{equation}
Now, the chirality condition becomes
\begin{eqnarray}
\omega &=& \gamma^{ab}\,\nabla_a \phi
\nabla_b \phi \nonumber =  \frac{1}{2\gamma g_{44}}\,[
(P^+ - \dot{\vec{X}}^2 )^2 - (X^{-'})^2]
(\vec{X}' \cdot \vec{X}') =0 \,,
\end{eqnarray}
from which is deduced that
\begin{equation}
X^{-'}= \pm (P^+ - \dot{\vec{X}}^2)\,.
\label{eq:solution}
\end{equation}
Plugging the Eq. (\ref{eq:solution}) in the constraint
(\ref{eq:constraint2}) we get the conditions that should
satisfy $\vec{a}$ and $\vec{b}$ in the chiral string
solution, namely,
\begin{equation}
\pm (P^+)^2 = (\vec{a}` + \vec{b}`)\cdot [ \vec{a}` 
- \vec{b}` \pm 
P^+ (\vec{a}` + \vec{b}`)]\,.
\label{eq:condition}
\end{equation}
Thus, this equation suggest to consider some cases. 

For instance, if we assume $P^+=1$, 
we get
\begin{equation}
1= 2(\vec{a}` + \vec{b}`)\cdot \vec{a}`  \,,
\end{equation}
which is different for the values for $\vec{a}$ and 
$\vec{b}$ reported in \cite{Vilenkin}. 

It is worthy to mention that the integrability
for the last cases was reached in absence of
electromagnetic field coupled to superconducting
strings. From the equations defining $X^+$, $X^-$ and the 
contraints (\ref{eq:constraint3}) and 
(\ref{eq:constraint4}) we can find the value for $\phi$. 
Again the vortons states are obtained when $\vec{b} `=0$ 
and  in such case $\vec{a}$ satisfy the relation 
$\pm P^{+2} = \vec{a}`\cdot [\vec{a} ` ( 1 \pm P^+)] $.

\section{Chiral Domain Wall Model}

In this section applying a similar mechanism
to that for conformal gauge chiral string model 
we get integrability for a chiral domain wall
model by means of a special ansatz. For
such intention, with all the former ingredients, 
we study a worldsheet described by the embedding
\cite{Vilenkin1},
\begin{equation}
X^{\bar{\mu}}(\tau,\xi^1,\xi^2)= 
\left(
\begin{array}{l}
\,\,\tau\\
\vec{\tilde{X}}
\end{array}
\right)\,\,,
\label{eq:dwembedd1}
\end{equation}
where
\begin{equation}
\vec{\tilde{X}}= 
\left(
\begin{array}{l}
\vec{{X}} \\
\phi
\end{array}
\right)\,.
\label{eq:dwembedd2}
\end{equation}
Now we can choose a gauge similar to the 
conformal gauge for strings , as follows 
\begin{equation}
\Gamma_{ab}= 
\left(
\begin{array}{ll}
\Gamma_{\tau \tau} & \,\, 0\\
\,0  & \Gamma_{AB}
\end{array}
\right)\,\,,
\label{eq:metric1}
\end{equation}
where $\Gamma_{AB} = g_{\bar{\mu} \bar{\nu}} 
X^{\bar{\mu}}{} _{,A}X^{\bar{\nu}}{}_{,B}$
and $\Gamma_{\tau A} =0$. We assume the special 
form for the embedding (\ref{eq:dwembedd1}),
as 
\begin{equation}
\vec{\tilde{X}}= \hat{\tilde{n}}\,\xi^2 + 
\vec{\tilde{X}}_{\perp} (\tau,\xi^1) \,,
\label{eq:dwgauge}
\end{equation}
with the conditions on $\hat{\tilde{n}}$ to be
a unit vector and perpendicular to 
$\vec{\tilde{X}}_{\perp}$. So, in this conformal 
gauge we have the constraints
\beq
\dot{\vec{\tilde{X}}}_{\perp} \cdot 
{\vec{\tilde{X}}} '_{\perp} &=& 0 \,,
\label{eq:const1} \\
\dot{\vec{\tilde{X}}}_{\perp} \cdot 
\dot{{\vec{\tilde{X}}}}_{\perp} + 
{\vec{\tilde{X}}}'_{\perp} \cdot 
{\vec{\tilde{X}}} '_{\perp} &=& 1 \,,
\label{eq:const2}
\eeq
and the condition $\sqrt{- \Gamma}\,\Gamma^{\tau \tau} 
= - 1$. It is straightforward to demonstrate that 
$\bar{\Gamma}:= {\mbox{det}}(\Gamma_{AB}) = - 
\Gamma_{\tau \tau}$. In this part of the work, $^.$ and
$'$ denote derivatives with respect to $\tau$ and $\xi^1$,
respectively.

Assuming $g_{\mu \nu}$ to be Minkowski's metric and resting in the
constraints (\ref{eq:const1}) and (\ref{eq:const2}), the
induced metric $\Gamma_{ab}$ takes the form,
\begin{equation}
\Gamma_{ab}= 
\left(
\begin{array}{lll}
-|\vec{\tilde{X}} _\perp '|^2 & \hspace{4mm} 0 & 0\\
\hspace{6mm} 0  & |\vec{\tilde{X}} _\perp '|^2 & 0 \\
\hspace{6mm} 0 & \hspace{4mm} 0 & 1
\end{array}
\right)\,.
\label{eq:dwmetric}
\end{equation}
According to the standard DNG equations of motion, in
our present case the corresponding ones are 
promoted as $\ddot{\vec{\tilde{X}}} _\perp  - 
\vec{\tilde{X}} '' _\perp =0$,
whose solutions have the form 
\begin{equation}
\vec{\tilde{X}} _\perp = \frac{1}{2}\vec{\tilde{a}} 
( t + \xi^1) + \frac{1}{2}\vec{\tilde{b}} 
( t - \xi^1) \,.
\label{eq:dwsolution}
\end{equation}
Imposition of the chirality for superconducting domain
walls, lead us to the relation
\be
|\vec{\tilde{X}}_\perp {}'|^2 (n^4)^2 = 
\dot{\phi}_\perp ^2 - {\phi}''_\perp {}^2 \,,
\label{eq:dwcondition}
\ee
where $n^4$ is the four component of the vector
$\hat{\tilde{n}}$. Furthermore, the constraints 
(\ref{eq:const1}) and 
(\ref{eq:const2})  read as $ |\vec{\tilde{a}}{}{}'|^2 
=1$ and $ |\vec{\tilde{b}}{}{}'|^2 =1$, or explicitly 
they are given by
\begin{eqnarray}
|\vec{a}{}'|^2  + g_{44} \phi ' {}^2 &=& 1\,, \\
|\vec{{b}}{}'|^2  + g_{44} \bar{\phi} ' {}^2 &=&1 \,,
\end{eqnarray}
where we have considered the notation $\vec{\tilde{a}} = 
(\vec{a} ,\,\, \phi)$ and $\vec{\tilde{b}} = 
(\vec{b} ,\,\, \bar{\phi})$, {\it i.e.,}
\begin{equation}
\vec{\tilde{X}}_\perp= 
\left(
\begin{array}{l}
\vec{{X}}_\perp = \frac{1}{2}\vec{{a}} ( t + \xi^1) + 
\frac{1}{2}\vec{b} ( t - \xi^1)
\\
\phi_\perp = \frac{1}{2}\phi( t + \xi^1) + 
\frac{1}{2}\bar{\phi}( t - \xi^1)
\end{array}
\right)\,.
\label{eq:dwvect1}
\end{equation}
Plugging (\ref{eq:dwvect1}) in the condition 
(\ref{eq:dwcondition}), the chirality condition is
expressed now as 
\be
\phi ` \,\bar{\phi} ` [1 + g_{44} (n^4)^2] = (n^4) ^2 
(1- \vec{a} ` \cdot \vec{b}` ) \,.
\ee
In a similar way as in the chiral string model case,
the last equation suggest some cases. For example,
the solution considering $n^4 =0$ and $\phi =0$ 
(or $\bar{\phi} =0$), correspond to a straight 
superconducting domain wall with a carrying current 
arbitrary cross section, but not including current 
along the $\xi^2$ direction.

\section{Conclusions}

In this work we have developed the dynamics of 
chiral membranes using geometrical techniques.
Using our description we can able of reproduce 
the results of \cite{Carter,Vilenkin,Davis1}
 in order to show consistency of our 
description. In fact, our scheme is resemble 
to DNG theory in five 
dimensions using a Kaluza-Klein approach. 
Integrability for both chiral string model and 
chiral domain wall model
was obtained using a simple ansatz. Besides of 
this description, the Hamiltonian analysis for the 
chiral model is part of an extense work \cite{ChiralMemb}. 
The full physical description is not over yet because 
a deep understanding of integration of equations 
of motion has not been accompplished. The search of new 
solutions for the case of chiral superconducting membranes 
is part of a forthcoming paper.

\section{Acknowledgements}

ER is indebted to R. Capovilla  for many valuable  
discussions and suggestions. The support in part from SNI 
(M\'exico) and CONACYT is grateful.

\end{document}